\documentstyle[12pt]{article}

\newcommand{\sect}[1]{\setcounter{equation}{0}\section{#1}\indent}
\renewcommand{\theequation}{\thesection.\arabic{equation}}
\begin{document}

\topmargin 0pt
\oddsidemargin 5mm
\def\bbox{{\,\lower0.9pt\vbox{\hrule \hbox{\vrule height 0.2 cm
\hskip 0.2 cm \vrule height 0.2 cm}\hrule}\,}}

\newcommand{\EQ}{\begin{equation}}
\newcommand{\EN}{\end{equation}}
\newcommand{\bea}{\begin{eqnarray}}
\newcommand{\ena}{\end{eqnarray}}
\newcommand{\hs}[1]{\hspace{#1 mm}}
\renewcommand{\a}{\alpha}
\renewcommand{\b}{\beta}
\renewcommand{\c}{\gamma}
\renewcommand{\d}{\delta}
\newcommand{\e}{\epsilon}
\newcommand{\shalf}{\frac{1}{2}}
\newcommand{\pa}{\partial}
\newcommand{\tri}{{\small $\triangle$}}
\newcommand{\dz}{\frac{dz}{2\pi i}}
\newcommand{\ra}{\rangle}
\newcommand{\lan}{\langle}
\newcommand{\nn}{\nonumber \\}
\def\a{\alpha}
\def\b{\beta}
\def\g{\gamma}
\def\G{\Gamma}
\def\d{\delta}
\def\D{\Delta}
\def\e{\epsilon}
\def\ve{\varepsilon}
\def\z{\zeta}
\def\t{\theta}
\def\vt{\vartheta}
\def\r{\rho}
\def\vr{\varrho}
\def\k{\kappa}
\def\l{\lambda}
\def\L{\Lambda}
\def\m{\mu}
\def\n{\nu}
\def\o{\omega}
\def\O{\Omega}
\def\s{\sigma}
\def\vs{\varsigma}
\def\S{\Sigma}
\def\vphi{\varphi}
\def\av#1{\langle#1\rangle}
\def\pa{\partial}
\def\na{\nabla}
\def\hg{\hat g}
\def\un{\underline}
\def\ov{\overline}
\def\cF{{\cal F}}
\def\Hsl{H \hskip-8pt /}
\def\Fsl{F \hskip-6pt /}
\def\cFsl{\cF \hskip-5pt /}
\def\ksl{k \hskip-6pt /}
\def\pasl{\pa \hskip-6pt /}
\def\tr{{\rm tr}}
\def\tcF{{\tilde{\cal F}}}
\def\tg{{\tilde g}}
\def\tB{{\tilde B}}
\def\cLNSI{{\cal L}^{\rm NS}_{\rm I}}
\def\bcLNSI{{\bar\cLNSI}}

\newcommand{\AP}[1]{Ann.\ Phys.\ {\bf #1}}
\newcommand{\NP}[1]{Nucl.\ Phys.\ {\bf #1}}
\newcommand{\PL}[1]{Phys.\ Lett.\ {\bf #1}}
\newcommand{\CMP}[1]{Comm.\ Math.\ Phys.\ {\bf #1}}
\newcommand{\PR}[1]{Phys.\ Rev.\ {\bf #1}}
\newcommand{\PRL}[1]{Phys.\ Rev.\ Lett.\ {\bf #1}}
\newcommand{\PTP}[1]{Prog.\ Theor.\ Phys.\ {\bf #1}}
\newcommand{\PTPS}[1]{Prog.\ Theor.\ Phys.\ Suppl.\ {\bf #1}}
\newcommand{\MPL}[1]{Mod.\ Phys.\ Lett.\ {\bf #1}}
\newcommand{\IJMP}[1]{Int.\ Jour.\ Mod.\ Phys.\ {\bf #1}}
\newcommand{\CQG}[1]{Class.\ Quant.\ Grav.\  {\bf #1}}
\newcommand{\PRep}[1]{Phys.\ Rep.\ {\bf #1}}
\newcommand{\RMP}[1]{Rev.\ Mod.\ Phys.{\bf #1}}

\begin{titlepage}
\setcounter{page}{0}

\begin{flushright}
COLO-HEP-368 \\
hep-th/9601167 \\
January 1996
\end{flushright}

\vspace{5 mm}
\begin{center}
{\large D-strings and F-strings from  string loops }
\vspace{10 mm}

{\large S. P. de Alwis\footnote{e-mail: dealwis@gopika.colorado.edu}
and K. Sato\footnote{e-mail: sato@haggis.colorado.edu}}\\
{\em Department of Physics, Box 390,
University of Colorado, Boulder, CO 80309}\\
\vspace{5 mm}
\end{center}
\vspace{10 mm}

\centerline{{\bf{Abstract}}}
Since the background fields of the string low energy action are
supposed to be the long range manifestation of a condensate of
strings, the addition of
world sheet actions to the low energy effective action needs some
string
theoretic explanation. In this paper we  suggest that this may be
understood,
as being due to string loop
effects. We
first present arguments using an equation due to Tseytlin and then
more
rigorously in the particular case of type IIB theory by invoking
the
Fischler-Susskind effect. The argument provides further
justification for
${\rm SL}(2,Z)$ duality between  D-strings and
F(fundamental)-strings. In an
appendix we
comment on recent attempts to relate the type IIA membrane to the 
11-dimensional membrane.

\vspace{29 mm}
\end{titlepage}
\newpage
\renewcommand{\thefootnote}{\arabic{footnote}}
\setcounter{footnote}{0}

\setcounter{equation}{0}
\sect{Introduction}

There has been  astonishing  progress in the last year  in
identifying and
elucidating connections between perturbatively different formulations
of string
theory.\footnote{It is hardly possible to list all the important
papers here so
we just mention one key paper which stimulated a lot of the
subsequent work
\cite{wit1}.} Nevertheless, an organizing (dynamical) principle that
would enable
one to understand this bewildering variety of connections is lacking.
In perturbative string theory
the relevant dynamical principle is world sheet Becchi-Rouet-Stora-Tyutin (BRST) (or conformal)
invariance. This requirement leads to equations of motion for the
background
fields. However, BRST invariance conditions, being short distance
effects on the
world sheet, will yield the same background equations independent of
topology, so
that the classical string background equations will be unmodified.
Clearly, this cannot be correct and the resolution was proposed by Fischler
and Susskind
\cite{fs}.  They pointed out that conformal anomalies can arise from
boundaries
of
 moduli space for genus higher than zero, and they showed that the
requirement
of BRST invariance for  the sum of contributions from the genus zero
and higher
order terms would modify the classical equations with stringy
quantum
corrections.
In this paper we will argue  that this dynamical principle has some
relevance
to the recent developments.

   String theory dualities require that classical solutions carrying
Ramond-Ramond (R-R) charges
be treated
on a par with elementary string states \cite{ht}. The former are
associated
with
non-perturbative effects while the latter are  the perturbative
spectrum of the
string which does not have any state carrying R-R charges. Nevertheless,
in
a recent
seminal paper it was shown by Polchinski \cite{jp} that the R-R
charge
carrying
classical solutions should be identified with the so-called D-branes,
i.e.,
branes to which open string ends are constrained to move on.
Polchinski's paper
opened up the possibility of extending the use of world sheet
methods to
formulate a systematic treatment of  non-perturbative effects.

In this paper we hope to make a small contribution to this program by
considering the effects of
R-R charge-carrying D-strings  in type IIB theory. Now, it is believed
that the
effect of having the D-brane is to add to the effective 10
dimensional action
the $(p+1)$-dimensional   action of the D-($p$-)brane. This
corresponds to the
addition of the fundamental string action to the effective action
by Dabholkar et al \cite{dghr}, in order to support the singular
string
solutions of the effective action. Indeed the type IIB string is
particularly
suited for the study of this correspondence since it is expected to
have
duality relating NS-NS fields and R-R fields in the effective action
\cite{ch},
\cite{js}. In   particular in the work of Schwarz \cite{js} a formula
for the
string tensions of strings carrying both NS-NS and R-R charges is
derived. These
strings were later interpreted by Witten \cite{wit2} as bound states
of  F(fundamental)-strings and D-strings.

 However, there is something strange about the addition of a world
sheet action
to the low energy effective action of string theory. This is usually
justified
in analogy with particle actions coupled to external fields.  But
unlike  the
case there, in string theory the background fields of the low energy
action are
not really external fields in which strings propagate---they are
expected to be
condensates of strings. At  resolutions larger than the string scale
the
stringy nature of the underlying reality will not be manifest. As one
approaches
the string scale the description in terms of  smooth fields (and
geometry)
should
break down. In particular, in regions of high curvature one expects
the
effective low energy
action description to be invalid, and presumably the singularity has  
no
physical
significance in string theory. The fact that the string coupling
vanishes as
one approaches the singularity\footnote{See, for example, the review of
Duff et
al \cite{dkl}.} is perhaps a reflection of this.

What we will seek in this paper  is a world sheet justification for
adding D-(and F-) string actions to the low energy effective actions.
Indeed,
 the
fact that these actions are (one or two) powers of $e^{\phi}$ down
from the
tree level effective action suggests that they should arise as loop
effects. We
will find that the Fischler-Susskind argument provides a rationale
correcting
the tree-level equations in this way. In particular, we will also
support from
the world sheet point of view the
$N=2$ supergravity argument for the   absence of dilaton couplings of  
R-R fields. The
modified Dirac-Born-Infeld (DBI) action that we get  is then shown to be of the same form
as that of the
fundamental string action
with tension given by Schwarz's formula.

The machinery that we use in this paper was developed in a remarkable
series
of papers by Callan, Lovelace, Nappi, and Yost (CLNY), culminating in
\cite{clny}. The
 construction of the D-brane state using their method was first done
by M. Li
\cite{ml}. During the course of this investigation, several papers
(\cite{ck},
\cite{md}, \cite{cs}) appeared that have some overlap with our work
(particularly the last paper). We feel, however, that none of these
has quite
addressed the issues from our perspective.  In particular, we give a
detailed
discussion of  BRST invariance in the presence of D-branes. In
Appendix A we
also illustrate
the
difference between our method of derivation of Schwarz's results for
bound
states of F- and D-strings, and that of Schmidhuber, by trying to
derive the
11D membrane action from the type IIA membrane. The latter can be
done only  in
the saddle-point
approximation. This is the case for  the passage  from the IIB
D-string to
F-string
as well that from
the IIA D-membrane to 11D membrane actions in \cite{cs}. In our case  
the former
is
exact. In Appendix B we discuss the relation between the ten  
dimensional
Born-Infeld action occurring in the work of CLNY \cite{clny} and the DBI
action of Leigh \cite{rl}.
\sect{The F-string action and SL(2,Z) duality in type IIB}
\indent

In the discussion of singular string (or more generally $p$-brane)
solutions of
the low energy action \cite{dghr}, it is usually assumed that the
singularity is
supported by the  explicit presence of a string. Thus, it is
interpreted as the solution to the equations of motion coming from
the original
effective action plus the world sheet action of the string, in
analogy with the
particle case. However, as mentioned in the introduction, there is a
crucial
difference between particle
motions
and that of the string. The statement that in  string theory the
background itself
is a condensate of strings may be summarized by Tseytlin's  equation
\cite{at}
for the
quantum string effective action.\footnote{This was used by Susskind
and Uglum
to provide support for
their
argument that  black hole entropy can be understood in terms of
strings \cite{su}.}
 This is a functional of
the
expectation value of the string field (whose low energy components
are the
metric, the dilaton, etc.) and is minimized with respect to
it (and indeed may make sense only at the minimum),
\EQ
\G_{10}=\int_{\chi=2} e^{-S_2[X]} +\sum_{\chi
=0,-2,...}\int_{\chi}e^{-S_2}\label{tseyt}.
\EN
In this equation the functional integral is taken over the embedding
functions
$X$ of the string world sheet in 10 dimensional  space-time and
intrinsic world
sheet metrics, divided by the volume of 2D diffeomorphism and Weyl
groups
weighted by the world sheet $\s$-model action,
\bea
S_2&=&-T_2\int d^2\s\, \biggl[{1\over 2}\sqrt{-\g}\g^{AB}\pa_A
X^{\m}\pa_B
X^{\n}
g_{\m\n}e^{\phi/2}+{1\over 2!}\e^{AB}\pa_A X^{\m}\pa_B X^{\n}
B_{\m\n}\nn
&&~~~~~~~~~~~~~~~-{1\over 4\pi T_2}\sqrt{-\g}R_{(2)}\phi \biggr],
\label{stringsigma}
\ena
(with the target space metric being the canonical one).
The sum is over the different world sheet topologies.\footnote{There
is a
subtlety
involving the M\"obius volume in the first term which we have ignored
since it
is  irrelevant to our argument.}
According
to Tseytlin, the first term is, in fact, the classical effective action
whose
bosonic part  (ignoring the RR sector) is given to
 leading order  in $\a '$  by
\EQ
I_{10}={1\over 2\k^2}\int d^{10}x\sqrt{-g}\,\left[R-{1\over 2}(\na
\phi)^2-{1\over 2\cdot 3!}e^{-\phi}H^2\right].
\label{3formsugra}
\EN
The quantum equations of motion are then given by,
\EQ
{\d\G\over\d\phi^i}={\d I_{10}\over\d\phi^i}-\sum_{\chi
=0,-2,...}\int_{\chi}e^{-S_2}{\d S_2
\over\d\phi^i}=0\label{quanteom}.
\EN
In the above the $\phi_i$ represent the different background fields $
G, B,
\phi,$ etc.
We will now argue that the terms coming from the addition of the 2D-string
action to the 10D-effective action in the work of \cite{dghr} and in
related
subsequent work\footnote{See \cite{dkl} and \cite{pt} for reviews.}
are in fact
obtained by approximating the leading string loop correction by its
sigma model
``classical" approximation.\footnote{For related observations, see \cite{at2}. We thank A. A. Tseytlin for bringing this reference to our attention.}

Just by using the condition that the string like solution that we are
looking
for preserves some supersymmetry one finds \cite{dghr},
\cite{dkl};
\EQ
ds^2=A_2^{-3/4}(y)\eta_{\a\b}dx^\a dx^\b+A_2^{1/4}(y)\d_{ij}dy^i
dy^j,
\label{fundstrmet}
\EN
$\a,\b=0,1; ~i, j=2,\ldots,9$ and
$e^{-\phi}=A_2^{1/2}(y)$, $B_{01}=-e^{2\phi}=-A_2^{-1}(y)$. We want
to
evaluate the
$\chi =0$ term in
(\ref{quanteom}) at its ``classical" point. So, in  addition to the
above we put
$\g_{AB}=\pa_A X^{\m}\pa_B X^{\n}g_{\m\n}$ and the ansatz
$X^{\a}=\s^{\a}, ~\a
=0,1;~~
X^i={\rm const.},~i =2,\ldots,9$, which gives a classical solution.
Substituting
into the action we find $S_2=0$ so that the leading  loop corrections
in
(\ref{quanteom})  are just what would come from adding the action
$S_2$
to the
effective action $I_{10}$.

Let us now  review Schwarz's results \cite{js}. In the bosonic sector
of type
IIB supergravity, there are a symmetric tensor $g_{\mu\nu}$, a
dilaton
$\phi$, and a 2-form gauge potential
$B^{(1)}_{\mu\nu}$ from the
NS-NS sector, and a scalar field $\chi$, another 2-form gauge
potential
$B^{(2)}_{\mu\nu}$, and a 4-form gauge potential $B_{\mu\nu\l\rho}$
from the R-R
sector. The
5-form field strength associated with $B_{\mu\nu\l\rho}$ is
self-dual, i.e.,
$F_5=*F_5$
where $F_5=dB_4$. It is impossible to write down an action when there
is such a
self-dual field strength. When $F_5=0$, however, one has (using the
canonical metric) the type IIB supergravity action in a manifestly
${\rm SL}(2,R)$-invariant form \cite{ch},
\EQ
I^{\rm IIB}_{10}={1\over 2\k^2}\int d^{10}x\sqrt{-g}\left[R
+{1\over 4}{\rm tr}(\pa {\cal M} \pa {\cal M}^{-1})
-{1\over 2\cdot 3!}{\cal H}^T {\cal M} {\cal H}\right].
\EN
${\cal M}$ is an ${\rm SL}(2,R)$ matrix of the scalar fields,
\EQ
{\cal M}=e^\phi \left(\matrix{|\l|^2&\chi\cr
                              \chi  &1   \cr}\right)
\in {\rm SL}(2,R).
\EN
$\l$ is a complex scalar field defined by $\l=\chi+ie^{-\phi}$. $\cal
H$ is a
vector of the two 3-form field strengths:
\EQ
{\cal H}=\left(\matrix{H^{(1)}\cr H^{(2)}\cr}\right).
\EN
Under an ${\rm SL}(2,R)$ transformation $\L$,
\EQ
\L=\left(\matrix{a&b\cr
                 c&d\cr}\right)
\in {\rm SL}(2,R),
\EN
${\cal M} \to \L {\cal M} \L^T$ or, equivalently, $\l \to
(a\l+b)/(c\l+d)$, and
${\cal H} \to (\L^T)^{-1} \cal H$.
 This symmetry implies that  the equations of motion yield a
multiplet of
(singular) string-like
solutions
carrying both NS-NS and R-R electric charges
\EQ
{\bf q} e_2={1\over \sqrt{2}\k}\int_{S^7} {\cal M}\,{}^*\!{\cal
H}.\EN
$e_2 = \sqrt2 \k T_2$
is the charge of the fundamental string with tension $T_2$ and  ${\bf
q}=
(q_1,q_2)^{T}$ is an ${\rm SL}(2,Z)$ vector of
integers  in accordance with the usual Dirac argument. The   tensions
(in
the
canonical metric) are given by $T =T_2\D_q^{1/2}$  and the second
factor is the
 ${\rm SL}(2,Z)$-invariant expression
\EQ \D_q={\bf q}^T {\cal M}_0^{-1} {\bf q} =
e^{\phi_0}(q_1-q_2\chi_0)^2+e^{-\phi_0}q_2^2.
\EN
The suffix ``0" denotes the v.e.v. of each field. The quantization of
charges implies that the
symmetry of the system breaks down from ${\rm SL}(2,R)$ to ${\rm
SL}(2,Z)$. The
solutions are of course
singular
 as is the case for the fundamental string. What one needs is a
string action
that
supports both the NS-NS and the R-R charges. The obvious
candidate is a string action (Nambu-Goto or Polyakov) with the
tension
$T$  and the two-form couplings
\EQ
-T_2\int d^2\s {1\over 2}\e^{AB}\pa_A X^\mu\pa_B X^\nu {\cal
B}^T_{\mu\nu}{\bf
q}.\label{Bcharge}
\EN
In our earlier considerations we justified the addition of such terms
for the
NS-NS charge  carrying (1,0) fundamental string as an approximate
evaluation of
 loop effect. In the rest of the paper we will
show how to obtain the same for strings with arbitrary ($q_1,q_2$)
charged
strings by what is essentially the same argument, but now extended to
world
sheets with boundaries coupled to D-branes (strings).
\sect{Tree-Level linearized equations of motion from BRST}
\indent

In this section we discuss (with the  modifications necessary  for
type IIB)
the
relevant parts of CLNY\cite{clny} (and \cite{fms}), which should be
consulted
for more
details.\footnote{In sections 3 and 4  we will work in a Euclidean
target
space. Also, in the rest of this paper we  use $T_2={1/2\pi\a'}=1$.}

The left-moving vertex operators are given by
$V^\mu_{-1}=\psi^\mu e^{-\varphi}c e^{ik\cdot X_{\rm L}}$,
$V^b_{-1}=2\pa\xi
e^{-2\varphi}c e^{ik\cdot X_{\rm L}}$, $V^c_{-1}={1\over2}\eta c
e^{ik\cdot X_{\rm L}},$
along with fermion vertex operator,
$V^A_{-1/2}=S^A e^{-\varphi/2}c e^{ik\cdot X_{\rm L}}$. (The
subscript on $V$
denotes the picture in which it is defined.)
We will also  need the  operator $U^B_{\eta}=-2^{-3/2}\eta S^B
e^{\varphi/2} c
e^{ik\cdot X_{\rm L}}$ later on. In the above $\psi^{\mu}$ is the
world sheet
superpartner of $X^{\mu}$ and $c,\varphi , \eta,\xi$ are ghost
fields. The
spin fields are defined by
$S^A=e^{(A\cdot\rho)}$,
where we have used the bosonization formula
$\psi^{\pm j}=e^{\pm\rho_j}, j=1,\ldots,5$.
$A$ is a spinor weight of ${\rm SO}(10)$ with five components each of
which
takes $\pm 1/2$ with
odd/even number of negative signs corresponding to chirality (dot/no
dot).

By tensoring left-moving vertex operators given above with the
corresponding
right-moving ones, we get the  $\s$-model interaction which describes
type IIB
superstring in the background of graviton $h$, dilaton $\mit\Phi$
(which we
have normalized differently from CLNY),
antisymmetric gauge potential $B^{(1)}$ from the NS-NS sector, and
three form field strength $H_3^{(2)}$ and one-form field strength
$H_1$ derived from scalar field $\chi$ from the R-R sector:
\bea
{\cal L}_{\rm I}&=&{1\over2}h_{\mu\nu}(x)V^{\{\mu}_{-1}{\tilde
V}^{\nu\}}_{-1}
-{\mit\Phi(x)}[V^b_{-1}{\tilde V}^c_{-1}-V^c_{-1}{\tilde
V}^b_{-1}]
+{1\over2}B^{(1)}_{\mu\nu}(x)V^{[\mu}_{-1}{\tilde V}^{\nu]}_{-1}
\nn
&+&{2^{-1/2}\over 3!}[{\Hsl}^{(2)}(x)C]_{AB}V^A_{-1/2}{\tilde
V}^B_{-1/2}
+{2^{-1/2}}[{\Hsl}_1(x)C]_{AB}V^A_{-1/2}{\tilde V}^B_{-1/2}.
\label{LI1}
\ena
In the expression above, the first three terms define ${\cal L}_{\rm I}^{\rm
NS}$, the
contribution from the NS-NS sector, and the rest define ${\cal
L}_{\rm I}^{\rm
R}$, that from the R-R sector. $C$ is the spinor metric.

The BRST charge $Q$ is given by
$Q=Q_0+Q_1+Q_2$
where
\EQ
Q_0=\oint{dz\over 2\pi i}e^\s T, \quad
Q_1=\oint{dz\over 2\pi i}\left({1\over2}i\eta\psi\cdot\pa X
e^\varphi\right),
\quad
Q_2=\oint{dz\over 2\pi i}\left(-{1\over4}e^\s\eta\pa\eta
e^{2\varphi}\right).\label{qbrst}
\EN
$T$ is the total stress tensor,
$T=-{1\over2}\pa X\cdot\pa X+{1\over2}\psi\cdot\pa\psi+{\rm ghosts}$.
The action of the  BRST operator on the left-moving vertex operators
gives
$[Q_0,V]_{\pm}={1\over2}k^2\pa c V$,
$~~[Q_2,V]_{\pm}=0$,
${}~~\{Q_1,V^\mu_{-1}\}=k_\mu V^c_{-1}$,
${}~~[Q_1,V^c_{-1}]=0$,
${}~~[Q_1,V^b_{-1}]=k_\mu V^\mu_{-1}$,
${}~~[Q_1,V^A_{-1/2}]=U^B_\eta (\ksl)_B{}^A$. These lead to
\EQ
[Q_0+{\tilde Q}_0, {\cal L}_{\rm I}]={1\over2}k^2{\cal L}_{\rm I}(\pa
c+{\bar\pa}{\tilde c}),
\label{Q0L}
\EN
\EQ
[Q_1+{\tilde Q}_1, {\cal L}^{\rm NS}_{\rm I}]
=-{i\over2}(\pa^\nu(g_{\nu\mu}+B_{\nu\mu}^{(1)})
+2\pa_\mu{\mit\Phi})[V^c_{-1}{\tilde V}^\mu_{-1}
-V^\mu_{-1}{\tilde V}^c_{-1}],
\label{Q1LNS}
\EN
\bea
[Q_1+{\tilde Q}_1, {\cal L}^{\rm R}_{\rm I}]
&=&2^{-1/2}[{1\over 3!}\g^{\l\mu\nu\rho}\g^{11}k_\l H^{(2)}_{\mu\nu\rho}
+{1\over 2}\g^{\nu\rho}k^\mu H^{(2)}_{\mu\nu\rho}
+\g^{\mu\nu}\g^{11}k_\mu H_\nu
+k^\mu H_\mu]_A{}^B
\nn
&&\times(U_\eta^A{\tilde V}_{B,-1/2}+V^A_{-1/2}{\tilde U}^\eta_B).
\label{2aA}
\ena
In deriving the last equation (which has been slightly modified
from CLNY for
future convenience), we used the chirality of  the vertex operators
$\g^{11}_{\,\,\,\,A}{}\!^B V_B=+V_A,
{}~~\g^{11}_{\,\,\,\,A}{}\!^B{\tilde U}_B=-{\tilde U}_A.$
{}~Antisymmetrized ~products ~of ~gamma ~matrices ~are ~defined ~by
$~~\g^{\mu_1\cdots\mu_n}=
{1\over n!}\sum_{\rm
perms}\e(p)\g^{\mu_{p(1)}}\cdots\g^{\mu_{p(n)}}$.

Linearized field equations are obtained from
\EQ
(Q+{\tilde Q}){\cal L}^{\rm NS}_{\rm I} |{\mit\O}\rangle=0.
\label{A}
\EN
$|{\mit\O}\ra$ is the $\rm SL_2$-invariant vacuum.
Thus, at tree level (linearized) we have, using (\ref{Q0L}):
\EQ
\bbox h_{\mu\nu}=0,\quad
\bbox B^{(1)}_{\mu\nu}=0,\quad
\bbox {\mit\Phi}=0,
\label{3A}
\EN
\EQ
\bbox H^{(2)}_{\mu\nu\l}=0,\quad
\bbox H_\mu=0.
\label{3B}
\EN
The gauge condition for the graviton is obtained from
eq.(\ref{Q1LNS})
\EQ
\pa^\nu(h_{\nu\mu}+B^{(1)}_{\nu\mu})+2\pa_\mu{\mit\Phi}=0.
\label{3C}
\EN
 From eq.(\ref{2aA}), we have
\EQ
dH^{(2)}=0,\quad
d\,{}^*\!H^{(2)}=0,\quad
dH_1=0,\quad
d\,{}^*\!H_1=0.
\label{3D}
\EN
We can generalize the argument given in \cite{clny} to obtain (the
massless
sector of) the
boundary
state in the presence of nonzero constant gauge field strength $F$
for D-branes
(strings). This is essentially given by Li \cite{ml} but we need some
details
which are not explicitly presented there. In NS-NS
sector this
state is given by
\bea
|B\ra_{\rm NS}&=&\int_{k_{\rm L},k_{\rm R}}
\k[\det(1+{\cal F})]^{1/2}
\bcLNSI
{1\over2}(\pa c+{\bar\pa}{\tilde c})(0)|{\mit\O}\rangle
\nn
&\equiv&\int_{k_{\rm L},k_{\rm R}}|k_{\rm L},k_{\rm R};{\rm
NS}\rangle
\label{4A}\\
&\equiv&\k\int_{k_{\rm L},k_{\rm R}}D_{\rm NS}(k_{\rm L},k_{\rm R})
{1\over2}(c_0+{\tilde c}_0)|{\mit\O}\rangle
\nonumber
\ena
with the definition
\EQ
\bcLNSI
=V^T_{-1}(k_{\rm L}){\underline T}{\tilde V}_{-1}(k_{\rm R})
+[V^b_{-1}(k_{\rm L}){\tilde V}^c_{-1}(k_{\rm R})-V^c_{-1}(k_{\rm
L}){\tilde
V}^b_{-1}(k_{\rm R})].\label{Lone}
\EN
$\k$ is the string coupling constant which later on we will set equal
to the
exponential of the dilaton. The definition of  $\cal F$ and the ${\rm
O}(10)$
rotation
matrix
$\underline T$ are given below.
The integral over left- and right-moving momenta is defined by
\EQ
\int_{k_{\rm L},k_{\rm R}}\equiv\int d^{10}k_{\rm L} d^{10}k_{\rm R}
\d^{p+1}(k_\parallel)
\d^{10}(k_{\rm L}-{\underline T}k_{\rm R}).
\label{measure}\EN
The first $\d$-function restricts the solution to the momentum  
eigenstates
with zero momentum in the parallel directions. The second  
$\d$-function
constraint is explained
below.

In the absence of gauge fields, we define $\underline T= T_0$ by
\EQ
T_0={\rm diag}[-{\bf 1}_{p+1},{\bf 1}_{9-p}],
\label{T0}
\EN
${\bf 1}_n$ is the $n\times n$ unit matrix.
Equations (\ref{measure}) and (\ref{T0}) imply free Neumann boundary  
condition on
$\a=0,\ldots,p$ and
free Dirichlet boundary condition on $i=p+1,\ldots,9$ as is
appropriate for a
D-($p$-)brane. Now, we turn on a gauge field coupled to the boundary
with
constant field strength  $\cF$. Taking into account the argument of
\cite{wit2}, we introduce
\EQ
{\cal F}=F+B^{(1)},\qquad {\cal F}^T=-{\cal F}.
\label{gifs}
\EN
   and write following \cite{clny}, \cite{ml},
\EQ
\underline{T} = T({\cal F})={1-{\cal F}\over 1+{\cal F}} T_0.
\label{T}
\EN
 Note
that $T$
is orthogonal because $\cal F$ is antisymmetric; $T^T T=TT^T={\bf
1}_{10}.$
   Now, we have the commutation relation
\EQ
[Q_1+{\tilde Q}_1, \bcLNSI]
=(k^\nu_{\rm L}T_{\nu\mu}-k_{\mu{\rm R}})V^c_{-1}(k_{\rm L}){\tilde
V}^\mu_{-1}(k_{\rm R})
-(T_{\mu\nu}k^\nu_{\rm R}-k_{\mu {\rm L}})V^\mu_{-1}(k_{\rm
L}){\tilde
V}^c_{-1}(k_{\rm R}).
\EN
BRST invariance requires that the r.h.s. of the equation above should
vanish.
Thus, we have
\EQ
k_{\rm L}^\nu T_{\nu\mu}-k_{\mu {\rm R}}=0,\qquad
T_{\mu\nu}k^\nu_{\rm R}-k_{\mu {\rm L}}=0.
\label{krotated}
\EN
This condition in fact requires that $\underline T$ is orthogonal.
We see that the $\d$-function constraint on eq.(\ref{4A}) ensures
$Q_1+{\tilde
Q}_1$ invariance.
Henceforth, we will be taking a static D-string
so that we
put
\EQ
{\cal F}=\left(\matrix{{\cal F}_{\a\b} & 0 \cr
                       0               & 0 \cr}
\right),\qquad\a,\b=0,1.\label{ftan}
\EN

The  boundary state in the R-R sector is expected to have the general
form (in the
$s_{\rm R}+s_{\rm L}=-1$ picture)
\EQ
|B\rangle_{\rm R}=\int_{k_{\rm L},k_{\rm R}}\sum_s V^A_s
L_{AB}{\tilde
V}^B_{-1-s}{c_0+{\tilde
c}_0\over
2}|{\mit\O}\ra,
\label{7A}
\EN
where the sum is over half integers $s$.
$L_{AB}$ is to be determined by using space-time supersymmetry.

The boundary state is expected to be supersymmetric under a linear
combination
of
the left and
right supersymmetry (SUSY) generators 
\EQ({\mit\L}_r^A+{\tilde {\mit\L}}_r^B
M_B{}^A(T))(|B\ra_{\rm NS}+|B\ra_{\rm R})=0,
\label{susystate}
\EN
where
$M_B{}^A(T)$ is the spinor representation of ${\rm O}(10)$ rotation;
i.e.,
\EQ
T_{\mu\nu}\g^\nu_A{}^B=[M(T)^{-1}\g_\mu M(T)]_A{}^B.
\label{7F}
\EN

In the case of D-string ($p=1$), one gets \cite{clny}, \cite{ml}
\EQ
[{\ksl}_{\rm R}]_A{}^B=[M(T)^{-1}{\ksl}_{\rm L} M(T)]_A{}^B,
\label{7G}
\EN
\EQ
M(T)=M\left({1-{\cal F}\over 1+{\cal F}}\right)M(T_0)
=[\det(1+\cF)]^{-1/2}e^{-{1\over2}{\cFsl}}(i\g_0\g_1).
\label{7H}
\EN
The action of the SUSY generators on the vertex operators is given in
a picture independent form by CLNY\cite{clny} (equation (3.37)). Using that   
and
(\ref{susystate}) one can determine \cite{ml}
 $L_{AB}$,
and  thus the R-R boundary state, which takes the form
\EQ
|B\ra_{\rm R}=\int_{k_{\rm L},k_{\rm R}} |k_{\rm L},k_{\rm R};{\rm
R}\ra,
\EN
\bea
|k_{\rm L},k_{\rm R};{\rm R}\rangle
&=&i{\k\over\sqrt{2}}\sum_s V_s^A[{\ksl}_{\rm L}
   e^{-{1\over2}{\cFsl}}(i\g_0\g_1)]_A{}^B
   {\tilde V}_{B,-1-s}{c_0+{\tilde c}_0\over2}|{\mit\O}\rangle
\nn
&\equiv&{\k\over\sqrt{2}}D_{\rm R}(k_{\rm L},k_{\rm R})
        {c_0+{\tilde c}_0\over2}|{\mit\O}\rangle.
\ena
This state is given as a sum over all the pictures which satisfies
$s_{\rm
R}+s_{\rm L}=-1$. For our calculations in the following sections, we
choose the
picture $s_{\rm R}=s_{\rm L}=-1/2$.\footnote{For further discussion
of this
point see \cite{clny}.}
\sect{Loop-Corrected Field Equations}
\indent
In this section we calculate the loop-corrected string field
equations in the
presence of a D-string. We need to
attach the boundary state obtained in the previous section to a
sphere using
the propagator
\cite{gm}, $\Pi\equiv{1\over2}(b_0+{\tilde b}_0)(L_0+{\tilde
L}_0)^{-1}$. So we
define the states
\EQ
|D\rangle_{\rm NS}\equiv\Pi|B\rangle_{\rm NS}=\int_{k_{\rm L},k_{\rm
R}}{\k\over 2k^2}D_{\rm NS}(k_{\rm L},k_{\rm R})|{\mit\O}\rangle,
\label{DNSstate}
\EN
\EQ
|D\rangle_{\rm R}\equiv\Pi|B\rangle_{\rm R}=\int_{k_{\rm L},k_{\rm
R}}{\k\over\sqrt{2}}{1\over 2k^2}D_{\rm R}(k_{\rm L},k_{\rm
R})|{\mit\O}\ra,
\label{DRstate}
\EN
where we have used the fact that only the massless modes have been
kept. The
important point here is that
even though the original boundary state is BRST invariant the state
modified by
the propagator is not.
Thus, this
BRST anomaly must be canceled by going offshell with the tree-level
equations.
The crucial point of \cite{fs} is that the modified
field equations should be obtained from the condition, $(Q+{\tilde
Q})|{\mit\Psi}\rangle=0$, where the state is defined  by adding
eqs.(\ref{DNSstate}) and (\ref{DRstate}) to the tree-level state
\EQ
|{\mit\Psi}\ra=({\cal L}^{\rm NS}_{\rm I}+{\cal L}^{\rm R}_{\rm
I})|{\mit\O}\ra
+\k\int_{k_{\rm L},k_{\rm R}} {1\over{2 k^2}}
\left( D_{\rm NS}+{1\over{\sqrt{2}}}D_{\rm R} \right)
|{\mit\O}\ra.
\EN
The condition, $(Q_0+{\tilde Q}_0)|{\mit\Psi}\rangle=0$,
gives the loop correction to the eqs.(\ref{3A}) and (\ref{3B}). In
the NS-NS
sector we obtain
\EQ
\bbox h_{\mu\nu}=\k T_{\{\mu\nu\}}\d^8(x_\perp)[\det(1+{\cal
F})]^{1/2},\label{heqn}
\EN
\EQ
\bbox B^{(1)}_{\mu\nu}=\k
T_{[\mu,\nu]}\d^8(x_\perp)[\det(1+\cF)]^{1/2},
\label{beqn}
\EN
\EQ
\bbox{\mit\Phi}=\k\d^8(x_\perp)[\det(1+\cF)]^{1/2},
\EN
and in the R-R sector
\EQ
\bbox H^{(2)}_{\mu\nu\l}=-{\k\over2}\pa_{[\l}J_{\mu\nu]},
\EN
\EQ
\bbox H_\mu={\k\over2}\cF^{\l\s}\pa_\mu J_{\l\s}.
\EN
$x_{\perp}$ denotes the 8-directions transverse to the surface of
D-string world sheet.
$J$ is a conserved 2-form current given by
\EQ
J={1\over2}\int d^2 \s{1\over\sqrt g}\d^{10}(f^\mu-x^\mu)\e^{AB}\pa_A
f^\l\pa_B
f^\s
g_{\l\mu}g_{\s\nu}dx^\mu\wedge dx^\nu.
\EN
We have introduced the D-brane embedding functions $f ^{\l}(\s )$,
$\s$ being
the D-sheet coordinates, and in our  calculation we had specialized
to
flat space and the static gauge  $f^\a=\d^\a_A\s^A,~f^i=0$.
The next condition
\EQ
(Q_1+{\tilde Q}_1)|{\mit\Psi}\rangle=0 \label{qone}
\EN
gives the loop correction to eqs. (\ref{3D}). No loop correction
to NS-NS
sector is given by this condition. The correction to R-R sector can
be read off
from
\EQ
(Q_1+{\tilde Q}_1)|D\ra_{\rm R}=-{\k\over2\sqrt2}\int_{k_{\rm
L},k_{\rm R}}
\left[U^A_\eta\tilde{V}_{B,-1/2}+V^A_{-1/2}\tilde{U}_{B,\eta}\right]
(\g^{\mu\nu}+\cF^{\mu\nu})_A{}^B{1\over2}J_{\mu\nu}|\mit{\O}\ra.
\EN
Thus, we get
\EQ
dH^{(2)}_3=0,\quad dH_1=0,\quad
\pa^\mu H^{(2)}_{\mu\nu\l}=i{\k\over2}J_{\nu\l},\quad
\pa^\mu H_\mu=i{\k\over4}\cF^{\nu\l}J_{\nu\l}.
\label{11A}
\EN
The last two equations in (\ref{11A}) can be written in form notation
\EQ
d\,{}^*\!H^{(2)}_3=i{\k\over2}\,{}^*\!J,\qquad
d\,{}^*\!H_1      =i{\k\over2}\cF\wedge\,{}^*\!J.
\label{twoform}
\EN
Now, recall that the current $J$ satisfies the  conservation equation
$d\,{}^*\!J=0$. The equations in (\ref{twoform}) are consistent with
it only
when $\k$ is constant. However, the coupling constant in string theory
is
actually
$\k=e^{\phi}$, where the ``curvature" dilaton $\phi$ is related to
the
``ghost'' dilaton ${\mit\Phi}$ by
\EQ
\phi(x)={\mit\Phi}(x)+{1\over 4}h^\mu{}_\mu(x).\label{dilaton}
\EN

 Thus,
as pointed out in \cite{clny},
we need to modify eq.(\ref{twoform}) to get
\EQ
d\,{}^*\!\left(e^{-\phi} H^{(2)}_3\right)={i\over2}\,{}^*\!J,\qquad
d\,{}^*\!\left(e^{-\phi} H_1\right)      ={i\over2}\cF\wedge{}^*\!J.
\label{mod2}
\EN
These modifications may be justified from our BRST point of view by
including
the extra contribution due to a linear dilaton in the BRST charge  
(see also
\cite{md}, \cite{ml2}).
When there
is a linear dilaton $\phi=X^0$ in the background the stress tensor is
\EQ
T_B=-{1\over2}\pa X\cdot\pa
X+{1\over2}\psi\cdot\pa\psi+\pa^2X^0.
\EN
It forms a supermultiplet with the superconformal current;
\EQ
T_F=-{1\over2}\psi\cdot\pa X+\pa\psi^0.
\EN
The second term on the r.h.s. modifies $Q_1$ in equation
(\ref{qbrst}). The commutators containing $Q_1$ thus get extra terms
\EQ
[Q_1,V^A_{-1/2}]=U^B_\eta \left\{
(\ksl)_B{}^A+i(\g^0)_B{}^A
\right\},
\EN
and
\bea
[Q_1+{\tilde Q}_1, {\cal L}^{\rm R}_{\rm I}]
&=&2^{-1/2}\Bigl[{1\over 3!}\g^{\l\mu\nu\rho}\g^{11}k_\l
H^{(2)}_{\mu\nu\rho}
+{1\over 2}\g^{\nu\rho}k^\mu H^{(2)}_{\mu\nu\rho}
+\g^{\mu\nu}\g^{11}k_\mu H_\nu
+k^\mu H_{\mu}
\nn
&&\quad+i\left\{{1\over 3!}\g^{0\mu\nu\rho}\g^{11}H^{(2)}_{\mu\nu\rho}
+{1\over 2}\g^{\nu\rho}\d^{0\mu}H^{(2)}{}_{\mu\nu\rho}
+\g^{0\nu}\g^{11}H_\nu
+\d^{0\mu}H_{\mu}\right\}\Bigr]_A{}^B
\nn
&&\quad\times(U_\eta^A{\tilde V}_{B,-1/2}+V^A_{-1/2}{\tilde
U}^\eta_B)
\nn
&=&2^{-1/2}(-i)\Bigl[
{1\over 3!}\g^{\l\mu\nu\rho}\g^{11}(\pa_\l-\pa_\l\phi)H^{(2)}_{\mu\nu\rho}
+{1\over 2}\g^{\nu\rho}(\pa^\mu-\pa^\mu\phi)H^{(2)}_{\mu\nu\rho}
\nn
&&\quad+\g^{\mu\nu}\g^{11}(\pa_\mu-\pa_\mu\phi)H_\nu
+(\pa^\mu -\pa^\mu\phi)H_{\mu}
\Bigr]_A{}^B
\nn
&&\quad\times(U_\eta^A{\tilde V}_{B,-1/2}+V^A_{-1/2}{\tilde
U}^\eta_B)
\nn
&=&\!\!-2^{-1/2}ie^{\phi}\Bigl[
{1\over 3!}\g^{\l\mu\nu\rho}\g^{11} \pa_\l\left(e^{-\phi}
H^{(2)}_{\mu\nu\rho}\right)
+{1\over 2}\g^{\nu\rho}\pa^\mu\left(e^{-\phi}
H^{(2)}_{\mu\nu\rho}\right)
\nn
&&\quad+\g^{\mu\nu}\g^{11}\pa_\mu\!\left(\!e^{-\phi}H_\nu\!\right)
\!+\pa^\mu\!\left(\!e^{-\phi} H_{\mu}\!\right)
\Bigr]\!_A{}\!^B
(U_\eta^A{\tilde V}_{B,-1/2}\!+\!V^A_{-1/2}{\tilde U}^\eta_B).
\ena
We used $k_\mu=-i\pa_\mu$, and $\d^0_\mu=\pa_\mu X^0=\pa_\mu\phi$.
Then from (\ref{qone}) we have the modified eq.(\ref{mod2}) and  
similarly
modified
Bianchi identities.

It is natural to define
\EQ
{\tilde H}^{(2)}_3\equiv e^{-\phi} H^{(2)}_3 ,\qquad
{\tilde H}_1      \equiv e^{-\phi} H_1.
\EN
The equations (\ref{mod2}) are now written as
\EQ
d\,{}^*\! {\tilde H}^{(2)}_3={i\over2}\,{}^*\!J,\qquad
d\,{}^*\! {\tilde H}_1      ={i\over2}\cF\wedge\,{}^*\!J,
\EN
and the  Bianchi identities  become
\EQ
d {\tilde H}^{(2)}_3=0,\qquad
d {\tilde H}_1      =0.
\EN
The effective action which gives these equations of motion and
Bianchi
identities should be in the form
\bea
S&\sim&{1\over2}\int_{M_{10}}
           \left( {\tilde H}^{(2)}_3\wedge{}^*\!{\tilde H}^{(2)}_3
                 +{\tilde H}_1\wedge{}^*\!{\tilde H}_1
                 +i\,{}^*\!J\wedge B^{(2)}
                 +i\chi\,{}^*\!J\wedge\cF
           \right)
\nn
&=&{1\over2}\int_{M_{10}}
           \left(  {\tilde H}^{(2)}_3\wedge{}^*\!{\tilde H}^{(2)}_3
                  +{\tilde H}_1\wedge{}^*\!{\tilde H}_1 \right)
  +{i\over2}\int_{\rm D-sheet} \left( \tB^{(2)}+\chi\tcF \right),
\label{effaction}
\ena
with $ {\tilde H}^{(2)}_3 = d B^{(2)},~{\tilde H}_1=d\chi$.
$\tB^{(2)}$ and $\tcF$ are the pullbacks of $B^{(2)}$ and $\cF$ to the D-sheet. Note that the redefined R-R fields appear without a dilaton factor.
\sect{From D-string action to F-string action}
\indent
The action which reproduces the right hand sides of the loop-corrected
equations  is obtained by adding the DBI action given by Leigh
\cite{rl} and the last two terms in
eq.(\ref{effaction});
\EQ
S_D=\int_{\rm D-sheet}
\left[e^{-\phi}\sqrt{\det(\tg+\tcF)}+i\tB^{(2)}
+i\chi \tcF\right].
\label{Dbraneaction}
\EN
$\tg$ is the pullback of $g$ to the D-sheet.
The only non-trivial equation of motion to  check is that for the
graviton
since, in particular, it is not  obvious  how the transverse
contributions to the
right hand side of  (\ref{heqn}) arise.
The point to note here is that the variation must be performed
keeping $\mit\Phi$
rather than $\phi$ fixed.  Putting
$g_{\mu\nu}=\d_{\mu\nu}+h_{\mu\nu}$, using
(\ref{dilaton}),  and going to the static gauge, we may write the
metric-dependent part of (\ref{Dbraneaction}) as
\bea
 S_D&\sim&\int d^{10}x\d^{8}(x_{\bot})e^{-{\mit\Phi}-{1\over
4}h_{\mu}^{\mu}}\sqrt {\det{}_{\|}  [1+h+{\cal F}]}\nn
&=& \int  
d^{10}x\d^{8}(x_{\bot})e^{-{\mit\Phi}}\sqrt{\det{}_{\|}[{1+{\cal
F}}]}\left[1+{1\over 4}
\tr_{\|} \big ({1-{\cal F}\over 1+{\cal F}}h\big )-{1\over
4}\tr_{\bot}h \right].\nn
\ena
It is worth noting that the sign change in the transverse directions
arises from the dilaton term where the transition from $\phi$ to
$\mit\Phi$ is the usual T-duality measure transformation.

We are now ready to show how this D-string action  becomes a
fundamental
string action with the tension given by Schwarz's formula. We need to  
start
with the action for $q_2$ D-strings so we take the following  
world sheet
Lagrangian\footnote{Note that from now on we will be working in a  
Minkowskian
signature metric space-time.}

\bea
{\cal L}&=&q_2\left\{e^{-\phi}\sqrt{-\det({\tilde g}+{\tilde{\cal
F}}})
+{1\over2}\chi\e^{\a\b}\tcF_{\a\b}+{1\over2}\e^{\a\b}\tB^{(2)}_{\a\b}
\right\}
\nn
&=&q_2 e^{-\phi}\sqrt{-\det({\tilde g}+{\tilde{\cal
F}}})+q_2\chi\tcF_{01}+q_2
\tB^{(2)}_{01}.
\label{lagrange}
\ena
Here, $\tcF_{01}={\dot A}_1-\pa_1 A_0-\tB^{(1)}_{01}$. We can rewrite
the DBI action (first term)
\bea
\sqrt{-\det(\tg+\tcF)}
&=&\sqrt{-\det\tg}\left[1-{1\over2}\tr
(\tg^{-1}\tcF\tg^{-1}\tcF)\right]^{1/2}
\nn
&=&\sqrt{-\det{\tilde
g}}\sqrt{1+(\tcF_{01})^2(\det{\tilde g})^{-1}}
\nn
&=&\sqrt{-\det\tg-(\tcF_{01})^2}.
\ena
The momentum conjugate to the gauge potential $A$ is
\EQ
\pi_1=-{q_2e^{-\phi}\tcF_{01}\over\sqrt{-\det{\tilde
g}-(\tcF_{01})^2}}+q_2\chi,
\qquad \pi_0=0,
\EN
which gives the Hamiltonian
\EQ
H=-\sqrt{-\det{\tilde g}}
\sqrt{q_2^2 e^{-2\phi}+(\pi_1-q_2\chi)^2}-A_0\pa_1\pi_1+\pa_1(\pi_1
A_0)+\pi_1
\tB^{(1)}_{01}+q_2 \tB^{(2)}_{01}.
\EN
We now choose to define the theory using the Hamiltonian form of the path integral.\footnote{This is of course not an unambiguous choice, since one may also define directly a Lagrangian path integral. However, our choice is the one which is directly related to the operator formulation and hence also to the argument of \cite{wit2}.}
After canceling the gauge group  
volume against $\int d\pi_0$, we have
\bea
Z&=&\int[d\pi_1][dA_0][dA_1]
\exp\left[{i\int d^2\s \left\{\pi_1{\dot
A}_1-H(A,\pi_1)\right\}}\right]
\nn
&=&\int[d\pi_1][dA_0][dA_1]
\exp\biggl[i\!\int\! d^2\s \Bigl\{-A_1{\dot \pi_1}+A_0\pa_1\pi_1
\nn
&&+\sqrt{-\det{\tilde g}}\sqrt{q_2^2e^{-2\phi}+(\pi_1-q_2\chi)^2}
-\pi_1 \tB^{(1)}_{01}-q_2 \tB^{(2)}_{01}\Bigr\}\biggr].
\ena
We can carry out the integrals over $A_0$ and $A_1$ to give
$\d$-functions
\bea
Z&=&\int[d\pi_1]\d({\dot\pi_1})\d(\pa_1\pi_1)
\nn&&\times
\exp\biggl[i\int d^2\s
\Bigl\{\sqrt{-\det{\tilde g}}\sqrt{q_2^2e^{-2\phi}+(\pi_1-q_2\chi)^2}
+\pi_1 \tB^{(1)}_{01}+q_2\tB^{(2)}_{01}\Bigr\}\biggr].
\ena
Because of the $\d$-functions, the integral reduces to the one over
the
zero-mode of $\pi_1$. The zero mode is quantized when $x^1$ is
compactified on
a circle \cite{wit2}. Thus the integral is replaced by a sum;
\EQ
Z=\sum_{q_1}\exp\left[i\int d^2\s
\left\{\sqrt{-\det{\tilde
g}}\sqrt{q_2^2e^{-2\phi}+(q_1-q_2\chi)^2}+q_1\tB^{(1)}_{01}+q_2
\tB^{(2)}_{01}\right\}\right].
\label{2BactionBq}
\EN
The combination $q_1\tB^{(1)}_{01}+q_2\tB^{(2)}_{01}$ is what we saw in
eq.(\ref{Bcharge}).
We can read off the string tension
\EQ
T=\sqrt{{q^2_2\over\k^2}+(q_1-q_2\chi)^2}.
\EN
In the canonical metric this expression is in the form $T\sim
\D_q^{1/2}$ given
in section 2.
These
facts support the ${\rm SL}(2,Z)$ invariance of the theory.

\sect{Discussion}
\indent
This action now provides the support for the singularity in the
solutions
generated by Schwarz \cite{js}. There are two issues remaining to be
discussed.
One is the fact that Leigh's action leads to string actions in the
Nambu-Goto rather than the Polyakov forms. This is easily understood
from our
earlier considerations that the string actions arise in this form
from the
classical $\s$-model approximation to the one-loop string term.  The
other is
the fact that  string theory calculations appear to be yielding
objects which
have space-time singularities. We believe that the resolution
of this
puzzle  comes from the following consideration. The string
like
solutions coming just from $I_{10}$ are such that $\k=e^{\phi}$ is
zero on the
singularity. Thus, the string action is relevant only for the
equations that
relate the charge to the string tension namely (\ref{mod2}) which
does not have
a vanishing
coupling constant factor on the right hand side.  This raises the
question
whether space-time singularities have physical  reality.
We should also mention here reference \cite{cpm}
where it is pointed out that the singularity becomes invisible to
strings when
the level-matching conditions are satisfied.\footnote{It should be
noted that
for a boundary state the BRST condition plus the boundary conditions
on the
ghosts imply the  level matching conditions $L_n-{\tilde L_{-n}}=0$  
as
well as
$F_n-{\tilde F_ {-n}}=0$.}

We believe that we have shed some light in this paper on the relation
between
the first quantized ($\s$- model) string and its counter part which
occurs as a
solution to effective low energy field equations. Polchinski's
observations
\cite{jp} hold out the promise that $p>1$ branes may also be analyzed
in terms
of world sheet considerations thus possibly obviating the necessity
for
the quantization of  $p>1$ actions. Perhaps some light on M theory
may also be
shed by
such considerations.
\vskip 0.5cm
\noindent
{\bf Acknowledgements:}\\
We  thank Joe Polchinski for correspondence. This work was
partially supported by Department of Energy contract No. DE-FG02-91-ER-40672.

\vskip 0.5cm
\appendix{\Large {\bf {Appendix A: Membrane Action}}}
\vskip 0.5cm
\renewcommand{\theequation}{A.\arabic{equation}}
\setcounter{equation}{0}
\indent
In ref.\cite{cs}, Nambu-Goto-type actions are obtained starting from
DBI
actions for both strings and membranes. This derivation relies on the
saddle-point approximation. As we saw in section 5, in the case of strings,
we can
obtain the same result without making any approximation. In this
appendix, we
apply the method used to derive eq.(\ref{2BactionBq}) to the case of
membrane.
It
turns out
that in this case we are unable to obtain the final result without
using a saddle point approximation, unlike in the case of the string.
We start
with the
effective
type IIA D-membrane action
\EQ
\int
d^3\s\left\{e^{-\phi}\sqrt{-\det(g+\cF)}+{1\over6}\e^{\a\b\g}
A_{\a\b\g}-{1\over2}\e^{\a\b\g}C_\a\cF_{\b\g}\right\}.
\EN
Here, $\a,\b,\g=0,1,2$. For simplicity, we consider the case where
the
space-time metric is Minkowskian. Then, we have
\bea
\sqrt{-\det(\eta+\cF)}&=&\left[1-{1\over2}
\tr(\eta^{-1}\cF\eta^{-1}\cF)\right]^{1/2}
\nn
&=&\sqrt{1-\cF_{01}^2-\cF_{02}^2+\cF_{12}^2}.
\ena
The canonical momenta are given by
\EQ
\pi_0=0,
\EN
\EQ
\pi_1=-e^{-\phi}\cF_{01}[1-\cF_{01}^2-\cF_{02}^2
+\cF_{12}^2]^{-1/2}-C_2,
\EN
\EQ
\pi_2=-e^{-\phi}\cF_{02}[1-\cF_{01}^2-\cF_{02}^2
+\cF_{12}^2]^{-1/2}+C_1,
\EN
and the Hamiltonian is
\bea
H=&-&\sqrt{1+\cF_{12}^2}\sqrt{e^{-2\phi}+(\pi_1+C_2)^2
+(\pi_2-C_1)^2}+C_0\cF_{12}+\pi_1\pa_1A_0+\pi_2\pa_2A_0
\nn
&-&{1\over6}\e^{\a\b\g}A_{\a\b\g}+\pi_1B_{01}+\pi_2B
_{02}.
\ena

Although we have gotten rid of the electric fields, the magnetic field
$\cF_{12}$
remains in the Hamiltonian; this is the main difference from the
D-string case.
The path integral goes almost the same way as that in the string case (again the
integral over $\pi_0$ is cancelled by the group volume):
\bea
Z&=&\int[d\pi_1][d\pi_2][dA_0][dA_1][dA_2]\exp\left[i\int
d^3\s\{\pi_1\dot{A}_1+\pi_2\dot{A}_2-H\}\right]
\nn
&=&\int[d\pi_1][d\pi_2][dA_0][dA_1][dA_2]
\nn
&&\times\exp\biggl[i\int
d^3\s\biggl\{(\pa_1\pi_1+\pa_2\pi_2)A_0+\pi_1\pa_0A_1+\pi_2\pa_0A_2
-C_0\cF_{12}
\nn
&&~~~~~~~~~~~~~~~~~~~~+\sqrt{1+\cF_{12}^2}\sqrt{e^{-2\phi}
+(\pi_1+C_2)^2+(\pi_2-C_1)^2}
\nn
&&~~~~~~~~~~~~~~~~~~~~+{1\over6}\e^{\a\b\g}A_{\a\b\g}-\pi_1
B_{01}-\pi_2B_{02}
\biggr\}\biggr]
\nn
&=&\int[d\pi_1][d\pi_2][dA_1][dA_2]\d(\pa_1\pi_1+\pa_2\pi_2)
\nn
&&\times\exp\biggl[i\int d^3\s\biggl\{
\pi_1\pa_0A_1+\pi_2\pa_0A_2-C_0\cF_{12}
\nn
&&~~~~~~~~~~~~~~~~~~~~+\sqrt{1+\cF_{12}^2}\sqrt{e^{-2\phi}
+(\pi_1+C_2)^2+(\pi_2-C_1)^2}
\nn
&&~~~~~~~~~~~~~~~~~~~~+{1\over6}\e^{\a\b\g}A_{\a\b\g}-\pi_1
B_{01}-\pi_2B_{02}
\biggr\}\biggr].
\ena
$\d$-function gives $\pi_1=-\pa_2y,~\pi_2=\pa_1y$ for a scalar
function $y$:
\bea
Z&\sim&\int
[dA_1][dA_2]\exp\biggl[i\int{d^3\s}\biggl\{\sqrt{1+\cF_{12}^2}
\sqrt{e^{-2\phi}+(\pa_1y-C_1)^2+(\pa_2y-C_2)^2}
\nn
&&~~~~~~~~~~~~~~~~~~~~~~~~+(\pa_0y-C_0)\cF_{12}
+{1\over6}\e^{\a\b\g}(A_{\a\b\g}+3\pa_{\a}yB_{\b\g})
\biggr\}\biggr].
\ena
It is impossible to integrate exactly over the magnetic field.
Taking a
variation of the action with respect to $\cF_{12}$ to find a saddle
point, we
get
\EQ
\cF_{12}=-{\pa_0y-C_0\over e^{-2\phi}+\eta^{\a\b}(\pa_\a
y-C_\a)(\pa_\b
y-C_\b)}.
\EN
$\cF_{12}$ can be eliminated in the action to get
\EQ
\int d^3\s \biggl\{\sqrt{e^{-2\phi}+\eta^{\a\b}(\pa_\a y-C_\a)(\pa_\b
y-C_\b)}
+{1\over6}\e^{\a\b\g}(A_{\a\b\g}+3\pa_{\a}yB_{\b\g})
\biggr\}.
\EN
This result can be generalized to general metric $g_{\a\b}$ and we
recover
Schmidhuber's derivation of the bosonic part of 11dimensional
supermembrane
action \cite{bst}.
\EQ
Z=\exp\left[i\int\left\{\sqrt{-\det{\hat
g}}+{1\over6}\e^{\a\b\g}{\hat
A}_{\a\b\g}\right\}\right],
\EN
where
\EQ
{\hat g}_{\a\b}=g_{\a\b}e^{-2\phi/3}+e^{4\phi/3}(\pa_\a
y-C_\a)(\pa_\b y-C_\b),
\EN
\EQ
{\hat A}_{\a\b\g}=A_{\a\b\g}+3\pa_\a y B_{\b\g}.
\EN
\appendix{\Large {\bf {Appendix B: The Born-Infeld action and Leigh's  
DBI
action}}}
\vskip 0.5cm
\renewcommand{\theequation}{B.\arabic{equation}}
\setcounter{equation}{0}
\indent
In this appendix we would like to discuss the relation between the  
Born-Infeld
action that comes as a prefactor in the CLNY \cite{clny}  
construction of the
NS-NS part of the boundary state equation (\ref{4A}) and the DBI  
action of
Leigh \cite{rl}. In the body of the text we considered a static  
D-string so
that the gauge field takes the form (\ref{ftan}) and then the  
relation is
trivial in the sense that the determinant of the ten-dimensional  
matrix
$1+{\cal F}$ is equal to the determinant of the two-dimensional   
matrix that
comes in Leigh's action in flat space and in static gauge. We will  
now show
this equivalence in the case that is  still restricted to flat space  
but now
allowing for general motions of the D-string.

\EQ
[\cF]_{\mu\nu}=\left(\matrix{\cF_{\a\b}&\cF_{\a j}\cr
                             \cF_{i\b} &\cF_{ij}  \cr}\right)
            =\left(\matrix{\cF_{\a\b}  & \pa_\a A_j
                                                                  \cr
                           -\pa_\b A_i
                                       & 0            }\right)
            =\left(\matrix{\cF  & Y\cr
                           -Y^T & 0\cr}\right),
\EN
where we defined
\EQ
Y_{\a j}\equiv \cF_{\a j}.
\EN
Recall that $\a$ and $\b$ are coordinates tangential to the  
D-$p$-brane
world
volume, while $i$ and $j$ are transverse to it, i.e.
$\a,\b=0,\ldots,p$, and $i,j=p+1,\ldots,9$. In the static gauge
\EQ
f^\a=\d^\a_A\s^A,
\EN
where $\s^A$ is the $p$-brane world volume coordinate.  Along with
the choice
of gauge potential,
\EQ
A_i=f^i(\s^A)=f^i(X^\a),
\label{Af}
\EN
the (flat space) DBI action is written as
\EQ
\det({\tilde g}+{\tcF})=\det(1+\cF+YY^T).
\EN
We expand the D-brane action up to fourth order in $A$ and $Y$
\bea
\ln\det(1+\cF+YY^T)
&=&{1\over2}\ln\det(1+\cF+YY^T)(1-\cF+YY^T)
\nn
&=&{1\over2}\tr\left(2YY^T-\cF^2-(YY^T)^2-{1\over2}\cF^4
+2\cF^2YY^T\right).
\label{DBI4}
\ena
The Born-Infeld action of CLNY  up to the same order is
\bea
\ln\det\left(\matrix{1+\cF  & Y\cr
                     -Y^T   & 1\cr}\right)
&=&{1\over2}\ln\det\left[\left(\matrix{1+\cF &Y \cr
                                       -Y^T  &1 \cr}\right)
                         \left(\matrix{1-\cF & -Y \cr
                                       Y^T & 1  \cr}\right)
\right]
\nn
&=&{1\over2}\ln\det\left\{1+\left(\matrix{YY^T-\cF^2 & -\cF Y \cr
                                          Y^T \cF    & Y^T Y
\cr}\right)\right\}
\nn
&=&{1\over2}\tr\left\{ 2YY^T-\cF^2-(YY^T)^2-{1\over2}\cF^4+2\cF^2YY^T
\right\}.
\label{CLNY4}
\ena
Equations.(\ref{DBI4}) and (\ref{CLNY4}) show that the DBI  action of Leigh  
and the
Born-Infeld action of CLNY
are equivalent up to fourth order in $A$ and $Y$. Since in the $p=1$  
case both
actions are of  this order
this is
sufficient to establish the equivalence for this case.
 These two actions are surely equivalent for any
$p$-brane but we have not tried to establish this in general.


\begin{thebibliography}{99}
\bibitem{wit1} E. Witten, \NP{B443}, 85 (1995), hep-th/9503124.
\bibitem{fs} W. Fischler and L. Susskind, \PL{B171}, 383 (1986);
\PL{B173}, 262 (1986).
\bibitem{ht}C. Hull and P. Townsend, \NP{B438}, 109 (1995), hep-th/9410167.
\bibitem{jp}J. Polchinski, \PRL{75}, 4724 (1995), hep-th/9510017.
\bibitem{dghr} A. Dabholkar, G. Gibbons, J. A. Harvey, F. Ruiz Ruiz,
\NP{B340}, 33 (1990); A. Dabholkar and J. A. Harvey, \PRL{63},
478 (1989).
\bibitem{ch} C. M. Hull, \PL{B357}, 545 (1995), hep-th/9506194.
\bibitem{js} J. H. Schwarz, \PL{B360}, 13 (1995), erratum, ibid. {\bf
B364}, 252 (1995), hep-th/9508143; hep-th/9509148; \PL{B367}, 97 (1996), hep-th/9510086.
\bibitem{wit2} E. Witten, \NP{B460}, 335 (1996), hep-th/9510135.
\bibitem{dkl} M. J. Duff, R. R. Khuri and J. X. Lu, \PRep{259}, 213
(1995), hep-th/9412184.
\bibitem{clny} C.G. Callan, C. Lovelace, C.R. Nappi and S.A. Yost,
\NP{B308}, 221 (1988).
\bibitem{ml} M. Li, \NP{B460}, 351 (1996), hep-th/9510161.
\bibitem{ck} C. G. Callan, Jr., and I. R. Klebanov, hep-th/9511173.
\bibitem{md} M. Douglas, hep-th/9512077.
\bibitem{cs} C. Schmidhuber, hep-th/9601003.
\bibitem{rl} R. G. Leigh, \MPL{A4}, 2767 (1989).
\bibitem{at} A. A. Tseytlin, \IJMP{A4}, 1257 (1989).
\bibitem{su} L. Susskind and J. Uglum, \PR{D50}, 2700 (1994), hep-th/9401070.
\bibitem{pt} P. K. Townsend, {\it Three Lectures on Supermembranes,
in Superstrings '88}, eds. M. Green, M. Grisaru, R. Iengo, E. Sezgin
and A. Strominger (World Scientific 1989).
\bibitem{at2} A. A. Tseytlin, \PL{B251}, 530 (1990).
\bibitem{fms} D. Friedan, E. Martinec and S. Shenker, \NP{B271}, 93
(1986).
\bibitem{gm} S. Giddings and E. Martinec, \NP{B278}, 91 (1986);
E. Martinec, \NP{B281}, 157 (1987).
\bibitem{ml2} M. Li, hep-th/9512042.
\bibitem{cpm} C. G. Callan, Jr., J. M. Maldacena, and A. W. Peet,
hep-th/9510134.
\bibitem{bst}E. Bergshoeff, E. Sezgin and P.K. Townsend, \PL{189B} 75
(1987);
\AP{185}, 330 (1988).
\end{thebibliography}
\end{document}